# Thermometry based on Coulomb-coupled quantum dots

Yanchao Zhang[a,*], Jincan Chen[b,**]

[a] *School of Science, Guangxi University of Science and Technology, Liuzhou 545006, People's Republic of China*

[b] *Department of Physics, Xiamen University, Xiamen 361005, People's Republic of China*

**Abstract**

A theoretical proposal that Coulomb-coupled quantum dots can be used as quantum probes to determine the temperature of a sample (i.e., an electronic reservoir) is proposed. Through the regulation of the positive or negative voltage bias in the thermometer, we are able to judge whether the temperature of the sample is higher or lower than that of the reference heat reservoir in the measure environment and to determine the precise temperature of the sample by using a particularly simple temperature-voltage bias relationship in the reversible condition. One outstanding characteristic of the thermometer is that when the sample is at low temperatures, a small temperature change will lead to a large voltage bias change. It means that the proposed thermometer has a high sensitivity when low-temperature samples are measured.

[*] Corresponding author.
[**] Corresponding author.
Email address: zhangyanchao@gxust.edu.cn (Y. Zhang), jcchen@xmu.edu.cn (J. Chen)

# 1. Introduction

Thermometry at the nanoscale has attracted considerable attention in the fields of modern science and technology because the conventional thermometry is not applicable when spatial resolution decreases to the submicron scale [1,2]. Precisely measuring the temperature of small systems at very low energies is an important issue in the development of nanotechnologies, which would pave the way towards many groundbreaking applications in quantum science [3,4], bioscience [5,6], and material science [7-12]. The urgent need for temperature measurement at the nanoscale motivated the development of precise quantum thermometric techniques. In recent years, various quantum systems have been investigated and demonstrated to be possibly used as thermometers. For example, a single quantum dot as a thermometer can be used to accurately estimate the temperature of fermionic [13,14] and bosonic [15,16] reservoirs. The smallest possible thermometer, namely, a single qubit, can distinguish two different temperatures of a bosonic bath [17]. The Coulomb blockade thermometer is also an important part of thermometry and has been studied for several decades [18-20]. It is based on the properties of the Coulomb blockade in tunnel junctions and the electronic temperature can be extracted from the current–voltage characteristics. Recent progress suggests that the unknown temperature of a sample can be estimated by putting it in thermal contact with an individual quantum probe, which can minimize the undesired disturbance on the sample [21]. A quantum thermal machine used to the low-temperature thermometry has been proposed in Ref. [22]. In this thermometry approach, a hot thermal reservoir and a cold sample are coupled to the machine to form a quantum refrigerator, which allows for simultaneously cooling the sample and determining its temperature in the case of the refrigerator reaching the Carnot efficiency [22].

A remarkable study based on Coulomb-coupled quantum dots was recently finished by Sánchez et al., who showed that the heat flow and the charge current can be completely decoupled by a three-terminal device, which consists of two capacitively coupled quantum dots connected to electron reservoirs operated in the Coulomb-blockade regime. They demonstrated that such a three-terminal quantum dot nano-scale heat engine can obtain the Carnot efficiency under reversible conditions [23]. In recent years, the electron and heat transport properties of Coulomb-coupled quantum-dot systems have been extensively investigated in the fields of thermoelectricity [23-27], thermal rectification [28-33], and logical stochastic resonance [34]. In this paper, we proposed that Coulomb-coupled quantum dots can be used as a quantum probe to determine the temperature of a sample through the transport properties in a three-terminal structure. It is expounded that the Coulomb-coupled quantum-dot thermometer can not only determine that the temperature of a sample is higher or lower than that of the reference heat reservoir in the measure environment but also precisely measure the temperature of the sample.

## 2. Model and theory

The model of a Coulomb-coupled quantum-dot thermometer is illustrated in Fig. 1. A tunneled quantum dot (denoted by $QD_B$) couples two separate electronic reservoirs with temperatures $T_L$ and $T_R$. A second quantum dot $QD_S$ is capacitively coupled to $QD_B$. Two quantum dots interact only though the long-range Coulomb force such that they can only exchange energy $U$ but no particles. The Coulomb-coupled quantum dots as a probe is tunneled to couple to a sample, which is an electronic reservoir with the temperature $T_S$ to be measured. The whole setup, apart from the sample, is referred to as a thermometer. The heat flow into the reservoir $\alpha$ ($\alpha = L, R, S$) is denoted by $J_\alpha$, and the charge current, in the thermometer, from the left reservoir to the right reservoir is denoted by $I$. The voltage bias $\Delta V$ is applied to the right reservoir and used to regulate the charge current.

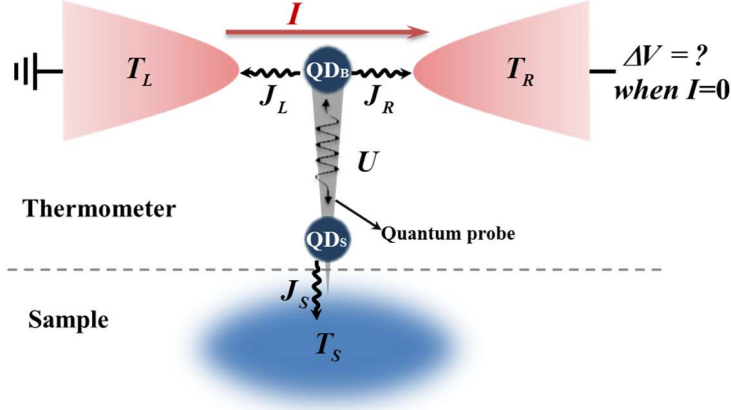

Fig. 1. The schematic illustration of a Coulomb-coupled quantum-dot thermometer.

In the Coulomb blockade regime, the quantum states of the Coulomb-coupled quantum-dot system are characterized by $|n_B n_S\rangle \in \{|00\rangle, |10\rangle, |01\rangle, |11\rangle\}$ with occupation probabilities $\boldsymbol{p} = (p_{00}, p_{10}, p_{01}, p_{11})$, where $n_B$ ($n_S$) is the occupation number of the quantum dot QD$_B$ (QD$_S$). In the sequential tunneling regime, the evolution of the occupation probabilities $\boldsymbol{p}$ can be described by a rate equation $\dot{\boldsymbol{p}} = \boldsymbol{M}\boldsymbol{p}$, where

$$M = \begin{pmatrix} -\Gamma_{L0}^+ - \Gamma_{R0}^+ - \Gamma_{S0}^+ & \Gamma_{L0}^- + \Gamma_{R0}^- & \Gamma_{S0}^- & 0 \\ \Gamma_{L0}^+ + \Gamma_{R0}^+ & -\Gamma_{L0}^- - \Gamma_{R0}^- - \Gamma_{S1}^+ & 0 & \Gamma_{S1}^- \\ \Gamma_{S0}^+ & 0 & -\Gamma_{L1}^+ - \Gamma_{R1}^+ - \Gamma_{S0}^- & \Gamma_{L1}^- + \Gamma_{R1}^- \\ 0 & \Gamma_{S1}^+ & \Gamma_{L1}^+ + \Gamma_{R1}^+ & -\Gamma_{L1}^- - \Gamma_{R1}^- - \Gamma_{S1}^- \end{pmatrix} \quad (1)$$

denotes the transition rates which are quite generally energy dependent. In Eq. (1), $\Gamma_{\alpha n}^{\pm} = \gamma_{\alpha n} f_\alpha^{\pm}(E_{B/S,n})$ describe tunneling events that take an electron into (+) or out (-) the quantum dot through electronic reservoir $\alpha$ when the occupation number the other quantum dot is $n$ ($n = 0,1$), $\gamma_{\alpha n}$ is the bare tunneling rate, $f_\alpha^+(x) = [1 + e^{(x-\mu_\alpha)/k_B T_\alpha}]^{-1}$ is the Fermi function and $f_\alpha^-(x) = 1 - f_\alpha^+(x)$, $E_{B/S,n} = \varepsilon_{B/S} + U_{B/S,0} + U\delta_{n1}$ is the charging energy of the quantum dot QD$_B$/QD$_S$, which depends on the occupation number $n$ ($n = 0,1$) of the other quantum dot, and $\varepsilon_{B/S}$ is the bare energy of the discrete level in the quantum dot QD$_B$/QD$_S$. The electrostaic energies are, respectively, given by [23]

$$U_{B0} = \frac{q}{C\tilde{C}}\left(\frac{q}{2}C_{\Sigma S} + C_{\Sigma S}\sum_{B=L,R}C_B V_B + CC_S V_S\right), \quad (2)$$

$$U_{S0} = \frac{q}{C\tilde{C}}\left(\frac{q}{2}C_{\Sigma B} + C_{\Sigma B}C_S V_S + C\sum_{B=L,R}C_B V_B\right), \quad (3)$$

$U_{B1} = U_{B0} + U$, and $U_{S1} = U_{S0} + U$, where $V_\alpha = \mu_\alpha/q$ is the electric potential of the reservoir $\alpha$, $\mu_\alpha$ is chemical potential of the reservoir $\alpha$, $q$ is the elementary charge, and $U = q^2/\tilde{C}$ is the exchanged energy between the two quantum dots when an electron tunnels into the empty quantum dot but leaves it only after the second electron has occupied the other quantum dot. The total capacitance of each quantum dot is defined by $C_{\Sigma B} = C_L + C_R + C$ and $C_{\Sigma S} = C_S + C$, and the effective capacitance $\tilde{C} = (C_{\Sigma B}C_{\Sigma S} - C^2)/C$.

We focus on the case of a nonequilibrium steady state. The steady-state probabilities are solved by the stationary solution of the rate equation, i.e., $\dot{\boldsymbol{p}} = 0$, together with the normalization condition $\sum \boldsymbol{p} = 1$. Thus, the charge current through the thermometer is given by

$$I = q\sum_n (\Gamma^-_{Rn} p_{1n} - \Gamma^+_{Rn} p_{0n}). \quad (4)$$

In the steady state, $I \equiv I_R = -I_L$. The heat flows from QD$_B$ into the left and the right reservoir are, respectively, given by

$$J_L = \sum_n (E_{Bn} - \mu_L)(\Gamma^-_{Ln} p_{1n} - \Gamma^+_{Ln} p_{0n}) \quad (5)$$

and

$$J_R = \sum_n (E_{Bn} - \mu_R)(\Gamma^-_{Rn} p_{1n} - \Gamma^+_{Rn} p_{0n}). \quad (6)$$

The heat flow from QD$_S$ into the sample reservoir is given by

$$J_S = \sum_n (E_{Sn} - \mu_S)(\Gamma^-_{Sn} p_{n1} - \Gamma^+_{Sn} p_{n0}). \quad (7)$$

This nonequilibrium steady state is characterized by a steady entropy flow given by

$$\sigma = \frac{J_L}{T_L} + \frac{J_R}{T_R} + \frac{J_S}{T_S}. \qquad (8)$$

### 3. Thermometry

In what follows, the thermometer is operated among the reference heat reservoir at temperature $T_B$ which is usually equal to the environment temperature and $T_L = T_R \equiv T_B$. The sample temperature $T_S$ is usually not equal to the temperature $T_B$ of the reference heat reservoir. The voltage bias $q\Delta V = \mu_R - \mu_L$ is applied to the thermometer, and the voltage bias and the temperature gradient can be used to precisely regulate the heat exchange process of the Coulomb-coupled quantum-dot system. The asymmetric energy-dependent transport barriers are defined by the bare tunneling rate $\gamma_{\alpha n} = \gamma$, except $\gamma_{L1} = \gamma_{R0} = \lambda\gamma$, where $0 \leq \lambda \leq 1$. Thus, the entropy flows as a function of the voltage bias are plotted in Fig. 2.

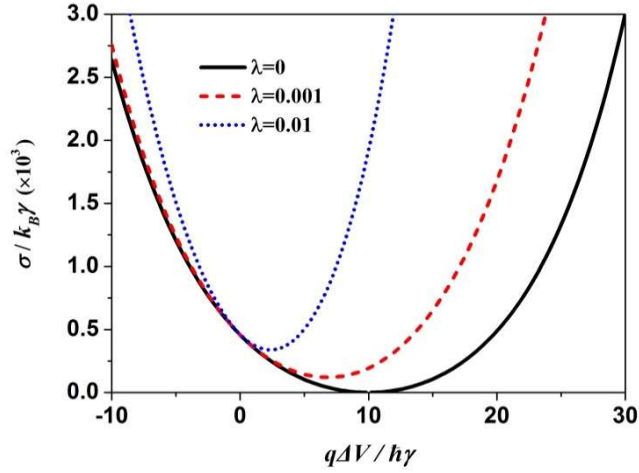

Fig. 2. Entropy flows as a function of the voltage bias for $\lambda = 0$ (black solid line), $\lambda = 0.001$ (red dashed line) and $\lambda = 0.01$ (blue dotted line). Parameters $k_B T_B = 5\hbar\gamma$, $k_B T_S = 10\hbar\gamma$, $q^2/C = 50\hbar\gamma$, $U = 20\hbar\gamma$, and $\varepsilon_B = \varepsilon_S = 0$ are chosen.

It is shown that in the nonequilibrium state ($\Delta T \neq 0$, $\Delta V \neq 0$), the entropy flows of the Coulomb-coupled quantum-dot system are always greater than zero, i.e., $\sigma > 0$, in the case of $\lambda \neq 0$. But, in the special case of $\lambda = 0$, i.e., $\gamma_{L1} = \gamma_{R0} = 0$, it can be found from Eqs. (5)-(8) that when the voltage bias and temperatures fulfill the following condition

$$q\Delta V = U\left(1 - \frac{T_B}{T_S}\right), \tag{9}$$

the entropy flow of the system is equal to zero, i.e., $\sigma = 0$. This result is also obtained in Ref. [35]. In this case, all the heat flows and charge current will vanish, so the system is in reversible in spite of the existence of the temperature gradient and voltage bias. This feature has been widely studied in thermoelectric devices [23,25]. Eq. (9) is a key ingredient for the proposed thermometer, which indicates an indirect measurement of the sample temperature by simply measuring the voltage bias in the reversible condition. Fig. 3 shows the charge current in the thermometer as a function of the voltage bias in the case of $\lambda = 0$ and sketches the scheme for measuring the temperature of the sample.

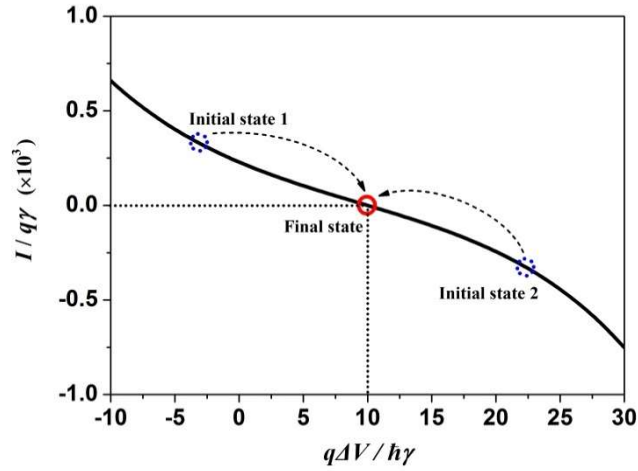

Fig. 3. The charge current in the thermometer as a function of the voltage bias for $\lambda = 0$. The other parameters are the same as those used in Fig.2.

The sample temperature can be determined by using the following strategy. First, there is a negative (or positive) charge current in the thermometer when the measurement starts, as shown by the blue dotted circle in Fig. 3. Then, increasing (or decreasing) the voltage bias and monitoring the charge current $I$ until the charge current is equal to zero, i.e., $I = 0$, we can obtain the value of the voltage bias at this time. Finally, we can determine the sample temperature $T_S$ by using Eq. (9). Fig. 4 shows the characteristic curve of the sample temperature and voltage bias in the reversible condition.

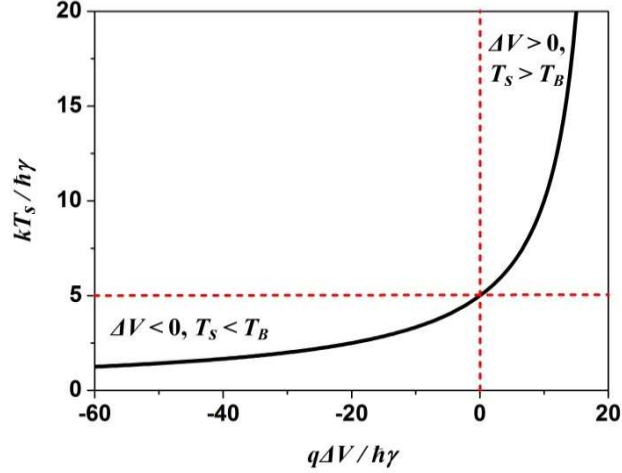

Fig. 4. The sample temperature as a function of the voltage bias in the reversible condition for fixed $k_B T_B = 5\hbar\gamma$ and $U = 20\hbar\gamma$.

It is found that when $\Delta V < 0$, the sample temperature is less than the reference temperature, i.e., $T_S < T_B$. However, in the region of $\Delta V > 0$, $T_S > T_B$. This suggests that the properties of the sample temperature (cold or hot relative to the reference temperature) can be judged by the value of the voltage bias without knowing the reference temperature of the measure environment. It can be also found that in the region of $\Delta V < 0, T_S < T_B$, a small change in the sample temperature can induce a large change in the voltage bias. This means that a low precision measurement of the voltage bias is converted into a high precision measurement of the sample temperature. This is characterized by a factor of the measurement sensitivity $\varphi$, which is defined as the change of the voltage bias divided by the variation of the sample temperature, i.e.,

$$\varphi = \frac{\partial \Delta V}{\partial T_S}. \qquad (10)$$

Fig. 5 shows the measurement sensitivity as a function of the sample temperature in the reversible condition. It can be found that when the sample temperature decreases, the sensitivity of measurement increases sharply. However, it should be pointed out that when the sample temperature decreases, the charge current in the thermometer will also decrease, so that it is more difficult to adjust the charge current to zero by the voltage bias. Therefore, when the thermometer proposed here is used to measure low temperature samples, the measurement difficulty increases with the decrease of the sample temperature.

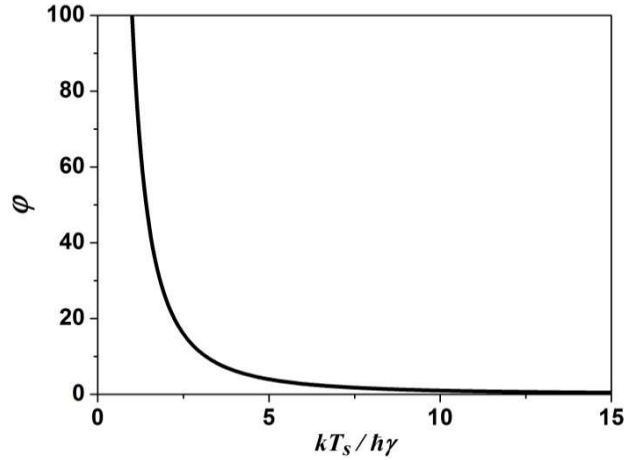

Fig. 5. Measurement sensitivity as a function of the sample temperature in the reversible condition for fixed $k_B T_B = 5\hbar\gamma$ and $U = 20\hbar\gamma$.

Recently, the Coulomb-coupled quantum dot system has been experimentally realized by growing a modulation doped GaAs/AlGaAs heterostructure with a high mobility two-dimensional electron gas (2DEG) located about 92 nm below the surface and the tunneling rate can be modulated around $\hbar\gamma \sim \mu$eV [24]. In this case, the environment temperature $T_B \approx 58.0$mK in this paper, we can found that the sample temperature $T_S \approx 14.5$mK be measured when the voltage bias $\Delta V \approx -0.06$mV and the measurement sensitivity $\varphi \approx 64$.

## 4. Conclusions

In conclusion, we have established a Coulomb-coupled quantum-dot thermometer in which the Coulomb-coupled quantum dots as a probe is tunneled to couple to a sample and the sample temperature can be determined through the temperature-voltage bias characteristics in the reversible condition. We also determined that the Coulomb-coupled quantum-dot thermometer can be used not only to directly judge the temperature of the sample to be higher or lower than that of the reference heat reservoir which is unnecessarily known but also to precisely measure the temperature of the sample. In principle, the proposed thermometer has a high sensitivity when low-temperature samples are measured. The results obtained here can provide theoretical guidance for the theoretical advance of temperature estimation in the quantum regime and open up potential technological applications for the development of high precision thermometry at the nanoscale.

**Acknowledgments**

We would like to thank Rafael Sánchez for helpful discussions. This paper is supported by the National Natural Science Foundation of China (No. 11675132), and by the Guangxi University of Science and Technology Foundation for PhDs (No. 18Z11).

**References**

[1] F. Giazotto, T. T. Heikkilä, A. Luukanen, A. M. Savin, J. P. Pekola, Rev. Mod. Phys. 78 (2006) 217.

[2] *Thermometry at the Nanoscale*, edited by L. D. Carlos and F. Palacio (The Royal Society of Chemistry, Cambridge, 2016).

[3] E. Martín-Martínez, A. Dragan, R. B. Mann, I. Fuentes, New J. Phys. 15 (2013) 053036.

[4] D. Halbertal, J. Cuppens, M. Ben Shalom, L. Embon, N. Shadmi, Y. Anahory, H. R. Naren, J. Sarkar, A. Uri, Y. Ronen, Y. Myasoedov, L. S. Levitov, E. Joselevich, A. K. Geim, E. Zeldov, Nature 539 (2016) 407.

[5] J. S. Donner, S. A. Thompson, M. P. Kreuzer, G. Baffou, R. Quidant, Nano Lett. 12 (2012) 2107.

[6] G. Kucsko, P. C. Maurer, N. Y. Yao, M. Kubo, H. J. Noh, P. K. Lo, H. Park, M. D. Lukin, Nature 500 (2013) 54.

[7] G.W. Walker, V. C. Sundar, C.M. Rudzinski, A W. Wun, M. G. Bawendi, D G. Nocera, Appl. Phys. Lett. 83 (2003) 3555.

[8] L. Spietz, K.W. Lehnert, I. Siddiqi, R. J. Schoelkopf, Science 300 (2003) 1929.

[9] P. Neumann, I. Jakobi, F. Dolde, C. Burk, R. Reuter, G. Waldherr, J. Honert, T. Wolf, A. Brunner, J. H. Shim, D. Suter, H. Sumiya, J. Isoya, J. Wrachtrup, Nano Lett. 13 (2013) 2738.

[10] O.-P. Saira, M. Zgirski, K. L. Viisanen, D. S. Golubev, J. P. Pekola, Phys. Rev. Applied 6 (2016) 024005.

[11] A. V. Feshchenko, L. Casparis, I. M. Khaymovich, D. Maradan, O.-P. Saira, M. Palma, M. Meschke, J. P. Pekola, D. M. Zumbühl, Phys. Rev. Applied 4 (2015) 034001.

[12] M. Zgirski, M. Foltyn, A. Savin, K. Norowski, M. Meschke, J. Pekola, Phys. Rev. Applied 10 (2018) 044068.

[13] F. Haupt, A. Imamoglu, M. Kroner, Phys. Rev. Applied 2 (2014) 024001.

[14] F. Seilmeier, M. Hauck, E. Schubert, G. J. Schinner, S. E. Beavan, A. Högele, Phys. Rev. Applied 2 (2014) 024002.

[15] C. Sabín, A. White, L. Hackermuller, I. Fuentes, Sci. Rep. 4 (2014) 6436.

[16] U. Marzolino, D. Braun, Phys. Rev. A 88 (2013) 063609.

[17] S. Jevtic, D. Newman, T. Rudolph, T. M. Stace, Phys. Rev. A 91 (2015) 012331.

[18] J. P. Pekola, K. P. Hirvi, J. P. Kauppinen, M. A. Paalanen, Phys. Rev. Lett. 73 (1994) 2903.

[19] T. Bergsten, T. Claeson, P. Delsing, Appl. Phys. Lett. 78 (2001) 1264.

[20] D. I. Bradley, R. E. George, D. Gunnarsson, R. P. Haley, H. Heikkinen, Y. A.

Pashkin, J. Penttilä, J. R. Prance, M. Prunnila, L. Roschier M. Sarsby, Nat. Commun. 7 (2016) 10455.

[21] L. A. Correa, M. Mehboudi, G. Adesso, A. Sanpera, Phys. Rev. Lett. 114 (2015) 220405.

[22] P. P. Hofer, J. B. Brask, M. Perarnau-Llobet, N. Brunner, Phys. Rev. Lett. 119 (2017) 090603.

[23] R. Sánchez, M. Büttiker, Phys. Rev. B 83 (2011) 085428.

[24] H. Thierschmann, R. Sánchez, B. Sothmann, F. Arnold, C. Heyn, W. Hansen, H. Buhmann, L. W. Molenkamp, Nat. Nanotechnol. 10 (2015) 854.

[25] Y. Zhang, G. Lin, J. Chen, Phys. Rev. E 91 (2015) 052118.

[26] P. A. Erdman, B. Bhandari, R. Fazio, J. P. Pekola, F. Taddei, Phys. Rev. B 98 (2018) 045433.

[27] R. S. Whitney, R. Sánchez, F. Haupt, J. Splettstoesser, Physica E 75 (2016) 257.

[28] T. Ruokola, T. Ojanen, Phys. Rev. B 83 (2011) 241404(R).

[29] F. Hartmann, P. Pfeffer, S. Höfling, M. Kamp, L. Worschech, Phys. Rev. Lett. 114 (2015) 146805.

[30] Y. Zhang, X. Zhang, Z. Ye, G. Lin, J. Chen, Appl. Phys. Lett. 110 (2017) 153501.

[31] R. Sánchez, H. Thierschmann, L. W. Molenkamp, Phys. Rev. B 95 (2017) 241401(R).

[32] Y. Zhang, Z. Yang, X. Zhang, B. Lin, G. Lin, J. Chen, Europhys. Lett. 122 (2018) 17002.

[33] T. Chen, X. Wang, Physica E 72 (2015) 58.

[34] P. Pfeffer, F. Hartmann, S. Höfling, M. Kamp, L. Worschech, Phys. Rev. Applied 4 (2015) 014011.

[35] R. Sánchez, M. Büttiker, Europhys. Lett. 100 (2012) 47008.